%% file: main.tex
\documentclass[reprint,superscriptaddress]{revtex4-2}

\usepackage[utf8]{inputenc}
\usepackage{amsmath}
\usepackage{amssymb}
\usepackage{graphicx}
\usepackage{import}
\usepackage{color}
\usepackage{tikz}
\usepackage{makecell}
\usepackage{booktabs}
\usepackage{placeins}
\usepackage{setspace}
\usepackage{parskip}
\usepackage{hyperref}
\usepackage[version=4]{mhchem}
\usepackage{chemmacros}
\chemsetup{
    modules = all,
    formula = mhchem
}
\usepackage{multirow}
\usepackage[capitalise]{cleveref}

\newcommand{\bnabla}{{\bm \nabla}}
\newcommand{\Dp}{\bm{\mathcal{D}}^{(\mathrm{e})}}
\newcommand{\Dps}{D^{(\mathrm{e})}}
\newcommand{\mup}{\mu_\mathrm{p}}
\newcommand{\mus}{\mu_\mathrm{s}}
\newcommand{\substset}[1]{\mathcal{S}({#1})}
\newcommand{\prodset}[1]{\mathcal{P}{(#1)}}
\newcommand{\catset}[1]{\mathcal{E}{(#1)}}
\newcommand{\st}{\bm{S}}
\newcommand{\stind}{\bm{S}_\text{ind}}
\newcommand{\ind}{\text{ind}}
\newcommand{\cchem}[1]{c_{#1}}
\newcommand{\ccat}[1]{\rho_{#1}}
\newcommand{\Cchem}{\bm{c}}
\newcommand{\Ccat}{\bm{\rho}}
\newcommand{\pertchem}{\delta \tilde{\Cchem}}
\newcommand{\pertcat}{\delta \tilde{\Ccat}}
\newcommand{\ofrt}{\left( \bm{r}, t \right)}
\newcommand{\ofrhc}{\left( \bm{c}, \bm{\rho} \right)}
\newcommand{\diag}[1]{\widehat{#1}}
\newcommand{\Dchem}{\bm{\mathcal{D}}^{(\mathrm{c})}}
\newcommand{\Dchems}{D^{(\mathrm{c})}}
\newcommand{\rhomat}{\diag{\bm{\rho}}}
\newcommand{\Jchem}{\left. { {\partial}_{\Cchem} \bm{v} } \right|_0}
\newcommand{\Jcat}{\left.  {\partial}_{\Ccat} \bm{v}  \right|_0}

\newcommand{\Rchem}{ \bm{S} \Jchem }
\newcommand{\Rchemind}{\left( \Rchem \right)_\text{ind}}
\newcommand{\Rcheminv}{\Rchemind^{-1}}
\newcommand{\Rcat}{ \bm{S} \Jcat }
\newcommand{\Rcatind}{ \left( \Rcat \right)_\text{ind} }
\newcommand{\nscc}{1,052,145 }
\newcommand{\nscconetofive}{5,137 }
\newcommand{\scale}[1]{{#1}_0}
\begin{document}

\title{Spatial self-organization of enzymes in complex reaction networks}

% Use letters for affiliations, numbers to show equal authorship (if applicable) and to indicate the corresponding author
\author{Vincent Ouazan-Reboul}
\affiliation{Max Planck Institute for Dynamics and Self-Organization, Am Fassberg 17, D-37077, G\"{o}ttingen, Germany}
\affiliation{LPTMS, CNRS-Université Paris-Sud, 91400, Orsay, France}
\author{Ramin Golestanian} 
\affiliation{Max Planck Institute for Dynamics and Self-Organization, Am Fassberg 17, D-37077, G\"{o}ttingen, Germany}
\affiliation{Rudolf Peierls Centre for Theoretical Physics, University of Oxford, OX1 3PU, Oxford, UK}
\author{Jaime Agudo-Canalejo}
\affiliation{Max Planck Institute for Dynamics and Self-Organization, Am Fassberg 17, D-37077, G\"{o}ttingen, Germany}
\affiliation{Department of Physics and Astronomy, University College London, WC1E 6BT, London, UK}

% Please include corresponding author, author contribution and author declaration information
% \authorcontributions{V.O-.R., R.G., and J.A-.C. designed the research, conducted the research, analyzed the data, and
% wrote the paper.}
% \authordeclaration{The authors declare no competing interests.}
% \correspondingauthor{\textsuperscript{1}To whom correspondence should be addressed. E-mail: ramin.golestanian@ds.mpg.de, j.agudo-canalejo@ucl.ac.uk}

\begin{abstract}
    % Max: 250 words
    % Current: 177 words
    Living systems contain intricate biochemical networks whose structure is closely related to their function and allows them to exhibit robust behavior in the presence of external stimuli. % essential to their functioning, 
    Such networks typically involve catalytic enzymes, which can have non-trivial transport properties, in particular chemotaxis-like directed motion along gradients of substrates and products.
    Here, we find that taking into account enzyme chemotaxis in models of catalyzed reaction networks can lead to their spatial self-organization in a process similar to biomolecular condensate formation.
    We develop a general theory for arbitrary reaction networks, and systematically study all closed unimolecular reaction networks involving up to six chemicals. Importantly, we find that network-wide propagation of concentration perturbations can be key to enabling self-organization.
    The ability to self-organize  is highly dependent on the relative signs of the chemotactic mobilities of the enzymes to their substrate and product and on the global network structure.
    We find that spontaneous self-organization through chemotaxis can provide an avenue for the self-regulation of metabolic activity in complex catalyzed reaction networks. % such as those in biological cells. 
    The network-induced interaction mechanism we uncover operates in the regime where the substrate molecules are diffusion-limited, suggesting that signaling molecules could take advantage of this scenario towards their functionality.
\end{abstract}

\maketitle

A universal feature of biological systems is their ability to robustly perform complex tasks as a result of elementary
stochastic processes, from performing the right biochemical reactions when and where needed to maintaining homeostasis
\cite{Phillips2012}. Many biochemical processes are performed or regulated by intricate biochemical networks whose
structure is intrinsically linked to their function \cite{newman2018Networks,tang2021Topology}. Examples include
circadian clocks \cite{vanzon2007Allosteric}, genetic regulation networks \cite{davidson2010Regulatory}, and metabolic
reaction networks \cite{jeong2000Largescale}. Biochemical reaction networks typically involve kinetically unfavorable
reactions that would not occur spontaneously in any reasonable time \cite{alberts2017Molecular},  and only take place
due to the action of catalytic enzymes that therefore drive the system out of (quasi-) equilibrium. Structural analysis
tools have previously been leveraged in order to understand the structure-function relationship of these networks
\cite{stelling2002Metabolic}, to help characterize their structure from experimental data
\cite{fell2018Metabolic,moreno-sanchez2008Metabolic,visser2002Mathematics}, and to design networks which can be
incorporated in synthetic biological systems \cite{alon2019Introduction}.

However, these and other studies of metabolic networks have so far relied on the so-called well-stirred  assumption,
i.e.~they assume that the concentrations of all the chemicals and enzymes in the system have spatially homogeneous
concentrations. In contrast, biological cells are spatially heterogeneous environments, as highlighted in particular by
the existence of biomolecular condensates: many proteins, including enzymes, aggregate into dense droplets at the right
place and at the right time \cite{banani2017Biomolecular,soding2020mechanisms,o2021role}. The precise mechanisms behind
the formation of enzyme-rich biomolecular condensates, also known as metabolons
\cite{sweetlove2018role,pedley2022Purinosome}, are not well understood. In particular, it is unclear to what extent they
can be fully described within equilibrium physics (e.g.~weak, multivalent enzyme-enzyme or enzyme-scaffold interactions)
or whether nonequilibrium effects due to enzyme catalytic activity play a role. Intriguingly, the composition of
metabolons is often finely tuned to the metabolic pathway at hand: all the enzyme species in a particular pathway are
present, in the right proportions so that the product of one enzyme species can be ``channeled'' as the substrate of the
next enzyme species in the pathway \cite{castellana2014enzyme,pareek2021metabolic,hinzpeter2019Regulation}. Moreover,
metabolons form and dissolve dynamically in response to the changing metabolic needs of the cell
\cite{sweetlove2018role,pedley2022Purinosome}. These observations, as well as in vitro experiments \cite{Testa2021} and
theoretical studies of model systems \cite{Kokkoorakunnel2024}, strongly suggest that nonequilibrium effects due to
catalysis play a role in the formation of metabolically active enzyme complexes, enabling a link between network
topology and spatial structure.

A minimal yet powerful hypothesis that could provide this link is that chemotactic-like motion of enzymes in the
presence of self-generated substrate and product gradients causes effective enzyme-enzyme interactions
\cite{agudo2018enhanced}, in analogy with chemotactic microbial aggregation \cite{keller1970initiation} and with the
self-organization of chemically-active colloids
\cite{golestanian2012,palacci2013living,schmidt2019light,Golestanian2022a}.
Indeed, many in vitro experiments have shown that biological enzymes
\cite{yu2009molecular,dey2014chemotactic,zhao2018Substratedriven,jee2018enzyme} can chemotax in response to gradients of
their substrate.
For the underlying mechanism, phoretic-hydrodynamic effects and binding-induced conformational changes of the enzyme
have been proposed as possible candidates \cite{agudo2018phoresis,agudo2018enhanced}.
Minimal theoretical models and simulations have shown that enzyme-enzyme aggregation through chemotaxis is indeed
possible, and can lead to metabolically optimized stoichiometries of the enzyme aggregate
\cite{agudo-canalejo2019Active,ouazan-reboul2021Nonequilibrium,ouazan-reboul2023Selforganization,ouazan2023network,ouazan-reboul2023Interaction,giunta2020cross,kocher2021nanoscale}.
One of the key nonequilibrium features of such interactions is non-reciprocity \cite{Soto2014}, which can lead to a
wealth of complex behavior in multicomponent condensates \cite{Parkavousi2025,Osat2022}. Importantly, these effective
enzyme-enzyme interactions due to chemotaxis can be made thermodynamically consistent and  arise generically in standard
models for the nonequilibrium (thermo)dynamics of multicomponent mixtures \cite{cotton2022CatalysisInduced}. However,
all studies so far have focused on very simple metabolic networks, considering only the interconversion between two
metabolites by one \cite{giunta2020cross,cotton2022CatalysisInduced} or several enzymes
\cite{agudo-canalejo2019Active,ouazan-reboul2021Nonequilibrium,kocher2021nanoscale}, or simple metabolic cycles without
branches \cite{ouazan-reboul2023Selforganization,ouazan2023network,ouazan-reboul2023Interaction}.

Motivated by these considerations, here we study the spatial self-organization of enzymes participating in arbitrarily
complex chemical reaction networks. We generalize the mathematical framework typically used for the structural analysis
of biochemical networks to account for spatial dynamics, including the chemotaxis of the enzymes in response to
gradients of metabolites (\cref{fig:act_cycles}), and analyze the stability of homogeneous (well-stirred) states to
spatial perturbations. We arrive to a general, concise, and readily applicable expression describing the spatial
stability of any metabolic network. We then focus our study on the set of all \nscc closed unimolecular reaction
networks involving up to six chemicals. For each of these networks, we systematically compare results obtained assuming
concentration-dependent rates (saturated regime in the Michaelis-Menten kinetics) to results obtained with
concentration-independent reaction rates (linear regime in the Michaelis-Menten kinetics). We find that, while in the
former case the emergent enzyme-enzyme interactions conform to our expectations, with enzymes interacting with one
another only if they chemotax in response to chemicals that the other produces or consumes, in the latter case
network-wide effects reflecting a far-ranging propagation of concentration perturbations throughout the catalytic
network can have a strong effect on the enzyme-enzyme interactions. These network-wide effects in turn affect the
self-organization of the system, and thus imply a sensitive dependence upon the full structure of the reaction network,
leading to a wide variation in the ability of different reaction networks to self-organize.

\section*{Results}
\subsection*{Catalyzed reaction network formalism}
\label{sec:model}

\begin{figure*}[hbtp]
    \centering
    \includegraphics[width=\textwidth]{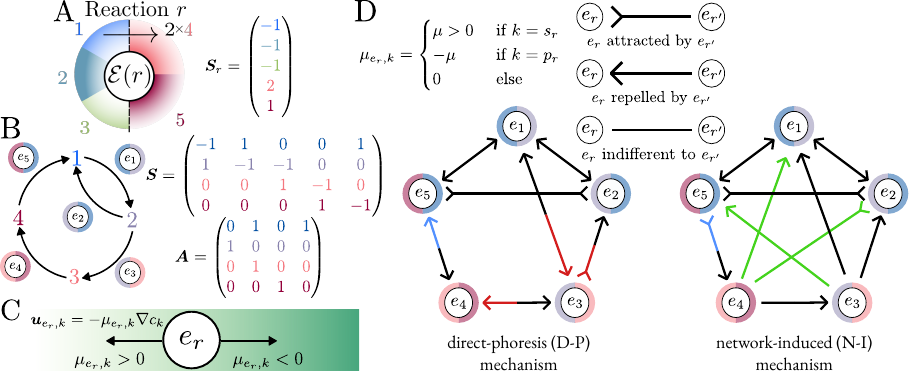}
    \caption{
    Definition of a catalyzed reaction network.
    (A) Reaction formalism: a given reaction $ r $ transforms a set of substrates $\substset{r}=\{1,2,3\}$ into a set of products $\prodset{r}=\{4,5\}$, catalyzed by a set of enzymes $ \catset{r} $.
        The reaction is associated to a stoichiometry vector $ \bm{S}_r $ describing the proportions of the production or
        consumption of the involved chemicals.
    (B) A set of reactions defines a chemical reaction network, whose structure is given by a stoichiometry matrix $
        \bm{S} $.
        The $ r\text{th} $ column of $ \bm{S} $ is the stoichiometry vector $ \bm{S}_r $. In this example, all reactions are unimolecular and each reaction $r$ is catalyzed by a different single enzyme $e_r$. The reaction network is thus equivalently described by an adjacency matrix $\bm{A}$.
    (C) The enzymes ($e_r$) involved in the reaction network are chemotactic, moving directionally in concentration
        gradients of the chemicals ($k$) which can take them towards high or low concentrations respectively for negative or positive chemotactic mobility $\mu_{e_r,k}$.
    (D) The combination of chemical activity and chemotaxis creates effective interactions between the enzymes,
        given by \cref{eq:intmat} and described by an interaction network. In this example, enzymes move
        towards higher (lower) concentrations of product (substrate). Saturated reaction kinetics (independent of substrate concentration) lead to interactions originating from {\it direct-phoresis} (D-P), where an enzyme only interacts with another if one has a chemotactic response to the substrates or products of the other. Non-saturated reaction kinetics (dependent on substrate concentration) create network-wide effects that can cause {\it network-induced} (N-I) interactions between enzymes that might even not interact in the D-P sense (green arrows), and can switch the direction of (blue half-arrows), or suppress (red half-arrows) the D-P mechanism.
}
    \label{fig:act_cycles}
\end{figure*}

We consider a minimal model for a generic catalyzed reaction network involving $ K $ chemicals, with concentrations
given by the space-and-time-dependent vector 
$ \Cchem\ofrt = \left(\cchem{1} \ofrt,  \cchem{2} \ofrt, \ldots  \cchem{K}
\ofrt\right)^\mathrm{T} $, which act as the substrates and products of $ M $ enzymes with concentrations 
$ \Ccat\ofrt = \left( \ccat{1} \ofrt, \ccat{2} \ofrt, \ldots, \ccat{M} \ofrt \right)^\mathrm{T}$
according to a set of $ R $ chemical reactions.
Each reaction $ r $ corresponds to the conversion of a set of substrate chemicals $ \substset{r} $ into a set of product
chemicals $ \prodset{r} $ assisted by a set of enzymes $ \catset{r} $, summarized as the chemical equation
\begin{equation}
    \label{eq:gen_chem_eq}
    \sum_{k_\mathrm{s} \in \substset{r} }^{} - S_{k_\mathrm{s}, r} \ k_\mathrm{s}
    \ce{ ->[\catset{r}] }
    \sum_{k_\mathrm{p} \in \prodset{r} }^{} S_{k_\mathrm{p}, r} \ k_\mathrm{p}.
\end{equation}
\Cref{eq:gen_chem_eq} features a set of stoichiometries $ \st_r = \left(S_{1,r}, S_{2,r}, \ldots,
S_{K,r}\right)^\mathrm{T} $, which give the number of molecules of each chemical $k$ that participates in reaction $ r
$: $ S_{k,r} $
is negative if $ k \in \substset{r} $, positive if $ k \in \prodset{r} $, and null otherwise
(\cref{fig:act_cycles}\textit{A}). The structure of the network formed by the set of $ R $ reactions is then defined by
the stoichiometry matrix $ \st \in \mathbb{Z}^{K \times R} $, whose $ r^\text{th} $ column is the stoichiometry vector
of reaction $ r $ (\cref{fig:act_cycles}\textit{B}).

In order to fully define the reaction network, we also need to specify the kinetics of its constitutive reactions, which
are given by a vector of reaction speeds  $ \bm{v}\ofrhc = \left( v_1\ofrhc, v_2\ofrhc, \ldots  v_R\ofrhc \right)^T $
whose $r^\text{th} $ component is the rate at which reaction $ r $ takes place. Taking into account the diffusion of
each chemical $ k $ with a coefficient $D^{(\mathrm{c})}_k$, the spatio-temporal dynamics of the chemical concentration
vector can then be written as
\begin{equation}
    \label{eq:evol_c_full}
    \partial_t \bm{c} \ofrt
    =
    \bm{S} \bm{v} \ofrhc
    +
    \Dchem \bnabla^2 \bm{c} \ofrt.
\end{equation}
Here, $ \mathcal{D}^{(\mathrm{c})}_{k,l} = D^{(\mathrm{c})}_k \delta_{kl} $ is the diagonal matrix of the diffusion
coefficients of the chemicals (where $\delta $ represents the Kronecker symbol) and $ \bnabla $ is the spatial gradient
operator. Note that Einstein's summation convention is not used throughout this work. \Cref{eq:evol_c_full} adds a
space-dependence to the reaction dynamics, which is commonly written in the context of well-stirred chemical reaction
networks \cite{blanchini2021Structural,feinberg2019Foundations,angeli2009Tutorial}, directly through the diffusion term
it contains and indirectly through the reaction speeds being a function of the space-dependent concentrations.

We describe the dynamics of the enzymes with a continuity equation that enforces the conservation of the total number of
enzymes. It involves a chemotactic drift term, which characterizes how a particle of enzyme species $m$ in a
concentration gradient of chemical $ k $ develops a velocity $ \bm{u}_{m,k} = - \mu_{m,k} \bnabla \cchem{k} $
(\cref{fig:act_cycles}\textit{C}).
$ \mu_{m,k} $ is the chemotactic mobility of enzyme species $ m $ in a concentration gradient of chemical $ k $. The set
of all mobilities defines the mobility matrix $ \bm{\mu} \in  \mathbb{R}^{M \times K} $.
This drift term is motivated by the observation of chemotactic motion in biological enzymes \cite{agudo2018enhanced}, as
discussed above. The evolution equation for the enzyme concentration vector is therefore
\begin{equation}
    \label{eq:evol_rho_full}
    \partial_t \bm{\rho} \ofrt = 
    \bnabla \cdot \left[
        \Dp \bnabla \bm{\rho} \ofrt +
        \rhomat \ofrt \bm{\mu} \, \bnabla \bm{c} \ofrt
    \right],
\end{equation}
where $ \rhomat_{m,n} = \ccat{m} \delta_{mn}$ and $ \mathcal{D}^{(\mathrm{e})}_{m,n} = D^{(\mathrm{e})}_m \delta_{mn} $
are diagonal matrices representing the enzyme concentrations and diffusion coefficients, which are introduced to help
with the brevity of the notation. 

\Cref{eq:evol_c_full,eq:evol_rho_full} describe a complex dynamics that couples the concentration fields of enzymes and
chemicals, and result in effective interactions between the $M$ enzyme species, mediated by the $K$ chemical
concentration fields (\cref{fig:act_cycles}\textit{D})

\cite{ouazan-reboul2021Nonequilibrium,agudo-canalejo2019Active,ouazan-reboul2023Selforganization,ouazan-reboul2023Interaction,yu2018Chemical,meredith2020Predator}.
The enzymes $ \catset{r} $ associated with the reaction $ r $ locally modify the concentrations of substrates $
\substset{r}$ and products $ \prodset{r} $ of $ r $, thus creating concentration gradients to which other enzymes can
respond through their chemotactic mobility. Throughout the rest of this work, we aim to determine if, and under which
conditions, the dynamics just described can lead to the spatial self-organization of the enzymes and chemicals
participating in a reaction network.

\subsection*{Homogeneous steady-state and linear stability analysis} \label{sec:lsa} In order to determine whether a
catalyzed reaction network undergoes self-organization, we perform a linear stability analysis on
\cref{eq:evol_c_full,eq:evol_rho_full}.
To do so, we perturb a homogeneous steady-state of the network. Because the enzyme concentrations are conserved, any
arbitrary homogeneous enzyme concentration vector $ \Ccat_0 $ is a valid steady-state. On the other hand, the
homogeneous chemical concentration vector $ \bm{c}_0 $ is defined by the requirement that the chemical reactions are
balanced, that is, must be a solution of the equation
\begin{equation}
    \label{eq:chem_HSS}
    \bm{S} \bm{v} \left( \bm{c}_0, \bm{\rho}_0 \right) = 0.
\end{equation}
In other words, the reaction rates associated with the homogeneous steady-state must be in the kernel $ \ker(\st) $ of
the stoichiometry matrix. The corresponding chemical concentration vector can then be obtained by deriving a vector of
rates $ \bm{v}_0 \in \ker(\st) $ and solving $ \bm{v} \left( \bm{c}_0, \bm{\rho}_0 \right) = \bm{v}_0 $ for $ \bm{c}_0
$.
The details of this calculation will depend on the network structure and choice of rate functions. In particular,
depending on whether or not the chemical reaction network satisfies local detailed balance, the homogeneous steady-state
will represent an equilibrium state or a nonequilibrium steady-state. In the equilibrium case, there will be a single
steady-state, and the concentrations $\Cchem_0$ will be independent of the choice of enzyme concentrations $\Ccat_0$. In
the nonequilibrium case, there may be more than one steady-state, and $\Cchem_0$ will generally depend on $\Ccat_0$. The
homogeneous concentrations obtained by solving \cref{eq:chem_HSS} are perturbed with space- and time-dependent
perturbations
% See notebook 5, p. 18-24, or notes in ~/These/Project_1.../3_random_networks/23_01_27
$ \Ccat \ofrt = \Ccat_0 + \delta \Ccat \ofrt $ and $ \Cchem \ofrt = \Cchem_0 + \delta \Cchem \ofrt $. Expanding
\cref{eq:evol_c_full,eq:evol_rho_full} to first order in the perturbations, and representing the perturbations as $
\delta \Cchem \ofrt = \pertchem(\bm{q}) \, e^{ i \bm{q} \cdot \bm{r} } \, e^{
\lambda(q^2) t } $ and $ \delta \Ccat \ofrt = \pertcat(\bm{q}) \, e^{ i \bm{q}  \cdot \bm{r} } \, e^{ \lambda(q^2) t }
$, yields the linearized equation
\begin{equation}
    \label{eq:full_stab_eq}
    \lambda (q^2)
    \begin{pmatrix} \pertcat \\  \pertchem \end{pmatrix}
    =
    \left(\begin{array}{c|c}
            - \Dp q^2 & - \rhomat_0 \bm{\mu} q^2  \\ \hline
            \st \Jcat & \st \Jchem - \Dchem q^2
    \end{array}\right)
    \begin{pmatrix} \pertcat \\  \pertchem \end{pmatrix},
\end{equation}
where
%$ \bm{I}_M $ is the $ M \times M $ identity matrix, and
$ \left( \Jcat \right)_{r,m}  =  {\partial}_{\ccat{m}} v_r | _{ \Cchem_0, \Ccat_0 } $ and $ \left( \Jchem \right)_{r,k}
= {\partial}_{\cchem{k}} v_r | _{ \Cchem_0, \Ccat_0 } $ are the Jacobians of the reaction rate vector.
\Cref{eq:full_stab_eq} is an eigenvalue problem that involves a matrix which couples the perturbation concentrations of
the enzymes and chemicals, whose eigenvalues $ \lambda(q^2) $ are the growth rates of perturbations associated with
wavevector $ \bm{q} $. It follows that if the matrix shown in \cref{eq:full_stab_eq} has at least one eigenvalue with
positive real part over some range of wavevectors, the corresponding system undergoes a self-organizing instability.

We observe that instabilities at the scale of the system size (corresponding to $ q \to 0 $ in \cref{eq:full_stab_eq})
occur for two different categories of modes: (i) $ \pertcat = 0 $ and $\lambda(0) \pertchem = \st \Jchem \pertchem$, or
(ii) $ \pertcat \neq 0 $ and $ \lambda(0) = 0 $. Case (i) corresponds to perturbations associated only with the
chemicals, which are not number conserving and originate from the well-stirred reaction dynamics. As we are interested
in the spatial self-organization of the full enzyme-chemical mixture, we assume that these chemical-associated modes are
stable, i.e. $ \bm{S} \Jchem $ has only negative eigenvalues, which is equivalent to assuming that the homogeneous
steady-state  $ ( \Cchem_0, \Ccat_0 ) $ is a stable fixed-point of the well-stirred reaction dynamics. Finding such
stable states is a well-studied problem in the theory of chemical reaction networks
\cite{clarke1980Stability,feinberg2019Foundations}, which we will not cover here. Similarly, we will assume that these
do not become positive at a finite band of $q^2>0$, as this would correspond to the well-known Turing instabilities of
the chemicals. Instead, in the following, we will focus on the modes of case (ii), which are associated with the enzymes
and respect number conservation.

\subsection*{Small wavevector expansion and effective enzyme-enzyme interactions}
For these enzyme-associated modes, the stability of the system is determined by the slope $ \lambda' ( 0 ) \equiv
\frac{d \lambda}{d q^2}\big\rvert_{q^2=0}$. If $\mathrm{Re}\left( \lambda' ( 0 )\right)>0$, the enzymes will display a
large-scale (but number-conserving) spatial instability.  In order to move forward, we make a quasi-static approximation
$ \partial_t \Cchem \simeq 0 $, equivalent to considering that the chemical concentration fields equilibrate very fast
relative to the timescale of variation of the enzyme concentration fields. The chemical part of the system in
\cref{eq:full_stab_eq}, i.e.~its last $K$ rows, then reduce to
\begin{equation}
    \label{eq:qss_perts}
    \left( \Rchem \right) \pertchem \simeq - \left( \Rcat \right) \pertcat + O(\Dchem \pertchem q^2),
\end{equation}
an equation relating the perturbations of the enzyme and chemical. To calculate the slope of the enzyme-related
eigenvalues at the origin, we must solve for $\pertchem$ in \cref{eq:qss_perts} and introduce it into the enzyme part of
the system in \cref{eq:full_stab_eq}, i.e.~its first $M$ rows. Note that the $O(q^2)$ term in \cref{eq:qss_perts} will
only lead to $O(q^4)$ terms in the eigenvalues, and thus can be ignored in the following.

Importantly, if the reaction network contains conserved moieties, i.e.~if at least one linear combination of the
chemical concentrations is conserved under the reaction dynamics, the matrix $ \st $ is not of full rank and inverting
\cref{eq:qss_perts} is not possible \cite{sauro1994Moietyconserved,reder1988Metabolic,cornish-bowden2002Role}. An
invertible equation can be derived by Gaussian elimination on $ \st $. This process yields a new stoichiometry matrix $
\stind \in \mathbb{Z}^{K_\ind \times R} $ associated with a set of $ K_\ind $ independent chemicals. As a byproduct of
this process, a set of $ K - K_\ind $ moiety conservation laws are derived, which can be written as $ \bm{C} \pertchem =
\bm{0} $ with $ \bm{C} \in \mathbb{Z}^{(K-K_\ind) \times K} $ a matrix of coefficients. We then build the Jacobian
matrices of the independent system, which have expressions
\begin{equation}
    \label{eq:jchemind}
    \Rchemind = 
    \begin{pmatrix}
        \stind \Jchem \\
        \bm{C}
    \end{pmatrix},
\end{equation}
and
\begin{equation}
    \label{eq:jcatind}
    \Rcatind = 
    \begin{pmatrix}
        \stind \Jcat \\
        \bm{0}
    \end{pmatrix}.
\end{equation}
The system $  \Rchemind  \pertchem = - \Rcatind  \pertcat $ is in general invertible
\cite{cornish-bowden2002Role,reder1988Metabolic} and gives solutions to  \cref{eq:qss_perts} with the appropriate moiety
conservation laws.

Plugging in $\pertchem = - \Rcheminv \Rcatind \pertcat$ into the enzyme part of \cref{eq:full_stab_eq}, we finally
obtain an eigenvalue equation for the slope $ \lambda' ( 0 )$, given by
\begin{equation}
    \label{eq:approx_stab_eq}
    \lambda'(0)   \pertcat
    = -  \left(  \rhomat_0 \bm{\eta} + \Dp \right)\pertcat,
\end{equation}
where we have defined the \emph{interaction matrix}
\begin{equation}
    \label{eq:intmat}
    \bm{\eta} \equiv - \bm{\mu} \Rcheminv \Rcatind
\end{equation}
with $ \bm{\eta} \in  \mathbb{R}^{M \times M} $. Equations~(\ref{eq:approx_stab_eq}) and (\ref{eq:intmat}) constitute a
central result of this work. Whenever $\mathrm{Re}(\lambda'(0))>0$, i.e.~if the matrix $  \rhomat_0 \bm{\eta} + \Dp$ has
at least one eigenvalue with negative real part, the system will display spontaneous spatial self-organization of the
enzymes. The interaction matrix has coefficients $ \eta_{m,n} $ representing the effective, chemical-field-mediated
response of enzyme species $ m $ to enzyme species $ n $. Note that $\bm{\eta}$ has the same physical dimensions as the
chemotactic mobility $\bm{\mu}$: effectively, $\bm{\eta}$ behaves as a chemotactic mobility in response to gradients of
enzymes, rather than gradients of chemicals. Negative (respectively positive) coefficients correspond to enzyme species
$ m $ being attracted to (respectively repelled by) species $ n $.

We note that the emergence of enzyme-enzyme interactions in this system is a nonequilibrium effect, caused by the lack
of detailed balance in the catalyzed chemical reaction network. Indeed, in an equilibrium state, the homogeneous
concentrations $\Cchem_0$ are independent of the enzyme concentrations, implying $\Rcat=0$ and therefore $\bm{\eta}=0$. 
Moreover, the interaction matrix $ \bm{\eta} $ is generally not symmetric. That is, the chemical-field-mediated
interactions are generally \emph{nonreciprocal}, so that the velocity of  enzyme $m$ in response to the presence of
enzyme $n$ is not necessarily of equal magnitude and opposite direction to that of $n$ in the presence of $m$, if $n$
and $m$ are not of the same type. In particular, $n$ may be attracted to $m$ while $m$ is repelled from $n$. Such
``chasing'' interactions \cite{Soto2014,Soto2015}, which would be impossible in an equilibrium system, are a
manifestation of the nonequilibrium origin of the effective enzyme-enzyme interactions, and are known to lead to
collective phenomena including travelling clusters and waves
\cite{agudo-canalejo2019Active,saha2020scalar,you2020nonreciprocity,Saha2025}.

There are two fundamentally different mechanisms with which enzymes can interact with each within this framework (see
\cref{fig:act_cycles}\textit{D}). Naturally, a given enzyme species develops an effective interaction with another
species if the latter produces or consumes chemicals that induce a chemotactic response in the former. We shall call
this mechanism {\it direct-phoresis} (D-P). However, this is not the only way that interactions can emerge in complex
chemical networks, as the dependence of $ \bm{v} $ on $ \bm{c}$ (as well as the existence of conserved moieties) can
induce a more complex structure in \cref{eq:intmat} through the screening factor $\Rcheminv$. This leads to the
emergence of a {\it network-induced} (N-I) mechanism, which can lead to non-intuitive behavior: for example, an enzyme
species $ m $ can interact with another enzyme species $ n $ despite having null chemotactic mobilities for the
substrates and products of $n$ (see \cref{fig:act_cycles}\textit{D}).

It is interesting to note that the D-P form of the interactions can be recovered in the special limit in which the
catalytic reactions take place in the saturated regime $ \bm{v}(\Ccat) $, i.e.~when they depend on the enzyme
concentrations but not on the chemical concentrations. In this special limit, an analogous calculation shows that the
stability of the system (away from the point of discontinuity $q^2=0$ for which $\lambda(0)=0$; see Refs.
\cite{agudo-canalejo2019Active,ouazan-reboul2021Nonequilibrium}) is governed by $\lambda(q^2)   \pertcat = -  \left( 
\rhomat_0 \bm{\eta}^\mathrm{sat} + \Dp q^2 \right)\pertcat$ with an interaction matrix $\bm{\eta}^\mathrm{sat} \equiv 
\bm{\mu} \, {\Dchem}^{-1} \Rcat$ which, noting that $\Dchem$ is a diagonal matrix, conforms to the picture of D-P
interactions.

\subsection*{Unimolecular reaction networks with bilinear rates} \label{sec:unimol_nets} We now apply the general
framework developed in the previous sections to a particular class of networks and reaction rate functions, to gain a
better understanding on how the properties of a catalytic network influence its stability.

We consider reaction networks containing only unimolecular reactions, i.e. a reaction $ r $ has a single substrate $ s_r
$ and product $ p_r $. Each column $ r $ of the stoichiometry matrix $ \bm{S} $ then only contains one $ -1 $ at line $
s_r $ and one $ +1 $ at row $ p_r $, the rest being only zeroes. Here, $s_r$ and $p_r$ are functions mapping reaction
indexes to chemical indexes. Moreover, every reaction is taken to be catalyzed by a single enzyme, so that $M=R$, and
the function $e_r$ mapping reaction indexes to enzyme indexes is a bijection (without loss of generality, one could
consider $e_r=r$). The structure of such unimolecular reaction networks can be characterized by an adjacency matrix $
\bm{A} \in \{0,1\}^{K\times K}$ defining a simple directed graph with $K$ nodes and $M$ edges, with $ A_{j,k} = 1 $
(corresponding to a directed edge) if a reaction converts chemical $ k $ into chemical $ j $, and 0 otherwise
(\cref{fig:act_cycles}\textit{B}).

To these networks, we associate bilinear reaction kinetics corresponding to enzymes operating in the linear
(non-saturated) concentration-dependence regime
\begin{equation}
    \label{eq:lowc_MM}
    v_r \ofrhc = \alpha_r \ccat{e_r} \cchem{s_r},
\end{equation}
where $ \alpha_r >0$ is an activity parameter corresponding, within a characteristic Michaelis-Menten picture, to the
ratio $k_\mathrm{cat}/K_M$ between the catalytic rate and the Michaelis constant of the enzyme $e_r$. Note that
\cref{eq:lowc_MM} represents unidirectional chemical reactions, with a negligible reverse rate, implying a strongly
out-of-equilibrium system. This could correspond e.g.~to a system where each catalyzed reaction is fuelled by the
consumption of a high-energy fuel such as ATP (in the biological context), or a light-fuelled photocatalytic system. 

\begin{figure}
    \centering \includegraphics[width=\columnwidth]{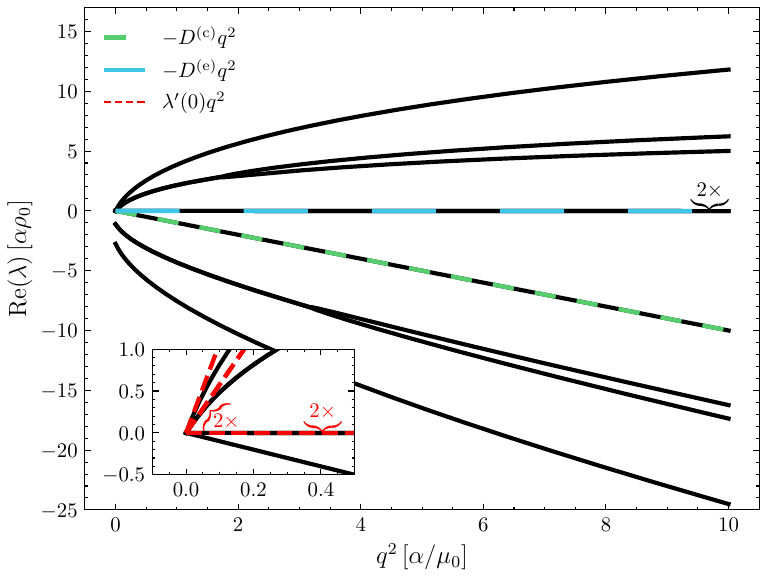} \caption{ Linear stability
        analysis for the unimolecular reaction network in \cref{fig:act_cycles}\textit{B} and the chemotactic mobilities
        in \cref{fig:act_cycles}\textit{D}, for linear reaction kinetics. The real parts of the eigenvalues for the full
        stability matrix \eqref{eq:full_stab_eq} are shown as solid black lines. Closed unimolecular reaction networks
        exhibit one purely diffusive chemical-associated diffusive mode (dashed green) and $ \max (1, M-K+1) $ purely
        diffusive enzyme-associated modes (dashed cyan). Inset: \cref{eq:approx_stab_eq} (dashed red) accurately
        captures the slope at the origin of the enzyme-associated eigenvalues of the full stability matrix. For this
    example, we chose $ \Dchems = \scale{\mu} c_\text{tot},  \Dps = \Dchems / 500, \mup = - \mus = 20$. } 
    \label{fig:ex_spectrum}
\end{figure}

Closed chemical reaction networks defined in this way have a unique steady-state with nontrivial dynamics only if the
associated adjacency matrix $ \bm{A} $ corresponds to a strongly connected directed graph, i.e.~if any chemical can be
converted into any other through some series of reactions.
If this condition is not satisfied, the reaction dynamics will lead to the accumulation of all the reaction material
into a subset of dead-end chemical species, with all the other chemical species having zero concentration. We thus focus
on these nontrivial networks, starting from the adjacency matrix
of a strongly connected digraph, and associating one chemical species to each of its nodes and a reaction together with
an associated enzyme to each of its directed edges (see \cref{fig:act_cycles}\textit{B} for an example of such a
network).
Networks generated according to this procedure always have a single homogeneous steady-state for the chemical
concentrations, $ \Cchem_0 $, which is stable under
space-independent, purely reactive dynamics according to the zero deficiency theorem \cite{feinberg2019Foundations}. As
the reaction rates in \cref{eq:lowc_MM} are linear in $ \cchem{s_r} $, finding the homogeneous concentration vector $
\Cchem_0 $ involves solving a linear system according to
\cref{eq:chem_HSS}. Because all transitions are unimolecular, there is automatically a moiety conservation corresponding
to the total concentration of chemicals, i.e.~$ c_\mathrm{tot} =  \sum_{k=1}^K c_{0,k}$ is conserved and is an input
parameter of the model. This conservation is also reflected in the fact that each column of the stoichiometry matrix
sums to zero.

\subsection*{Stability of networks with uniform parameters}  

To further reduce the parameter space and focus on the role of network topology, we make a number of simplifying assumptions in all that follows. We take $\alpha_r=\alpha$ to be equal for all enzymes, and choose all enzymes to have identical homogeneous concentrations $ \rho_{0m} =
\rho_0 $. We further assume that each enzyme species only chemotaxes in concentration gradients of its substrate and
product, and that the mobilities towards substrate and product are equal for all enzyme species, implying that the only non-null coefficients of the mobility matrix are of the form $ \mu_{e_r,
s_r} = \mus \scale{\mu} \ $ and $ \mu_{e_r,p_r} = \mup \scale{\mu} $, where $ \mus
$ and $ \mup $ are dimensionless parameters and $ \scale{\mu} >0$ is a mobility scale.
Finally, we take the same diffusion coefficients $\Dchems$ and $\Dps$ equal for all chemical and enzyme species, respectively.

With these simplifying assumptions, the eigenvalue equation for $\lambda'(0)$ in \eqref{eq:approx_stab_eq} can be
rewritten as
\begin{equation}
    \label{eq:approx_stab_eq_2}
    \beta   \pertcat
    = -  \tilde{\bm{\eta}} \pertcat,
\end{equation}
where $\lambda'(0)=\mu_0 c_\mathrm{tot} \beta-\Dps$, and $\tilde{\bm{\eta}} \equiv (\rho_0/c_\mathrm{tot})
(\bm{\eta}/\mu_0)$  is a rescaled, dimensionless form of the interaction matrix that only depends on $\mus$, $\mup$, and
the topology of the reaction network. Importantly, these parameters are sufficient to determine whether a given reaction
network has the potential to display self-organization or not. Indeed, let us focus on the eigenvalue with largest real
part, and define $\beta_\mathrm{max}\equiv\text{max}(\text{Re}(\beta))$. If $\beta_\mathrm{max}\leq 0$, the network is
guaranteed to be stable (as this implies $\mathrm{Re}(\lambda'(0))<0$). On the other hand, if $\beta_\mathrm{max}>0$,
such a network will display a spatial instability whenever the condition
\begin{equation}
    \frac{\Dps}{\mu_0 c_\mathrm{tot}} < \beta_\mathrm{max}
    \label{eq:beta_inst}
\end{equation}
is satisfied, as this implies $\mathrm{Re}(\lambda'(0))>0$. Thus, in systems with $\beta_\mathrm{max}>0$, an instability
can occur if the chemotactic mobility scale or total chemical concentration are sufficiently large, or if the enzyme
diffusion coefficient is sufficiently small.

To test the validity of \cref{eq:approx_stab_eq_2} and its general form \cref{eq:approx_stab_eq}, we have compared its
predictions for the slope $\lambda'(0)$ of the $M$ enzyme-associated eigenvalues, to the $M+K$ enzyme- and
chemical-associated eigenvalues $\lambda(q^2)$ calculated from the full problem, \eqref{eq:full_stab_eq}, for a variety
of networks and choices of parameters. The results for an example network are shown in \cref{fig:ex_spectrum}. Because
of the conservation of the total chemical concentration, we can expect one purely diffusive chemical-associated mode
going as $ - \Dchems q^2 $, which we observe (\cref{fig:ex_spectrum}, dashed cyan line). Similar purely diffusive modes
going as $ - \Dps q^2 $ are observed for the enzymes. Our numerical calculations show that there is one such mode if $ M
\leq K $, and $ M - K + 1 $ such modes if $ M > K $ (\cref{fig:ex_spectrum},  dashed green line).
\cref{eq:approx_stab_eq_2} always correctly captures the slope at the origin of  the $M$ enzyme-associated modes
(\cref{fig:ex_spectrum}, inset, dashed red lines). Importantly, of all $M+K$ modes, only the enzyme-associated ones can
become unstable, and the instability always occurs at $q^2 \to 0$. Overall, comparison with the numerical solution of
the full eigenvalue problem in \eqref{eq:full_stab_eq} confirms that, in order to study the stability of a unimolecular
reaction network with uniform parameters, it is sufficient to consider the reduced eigenvalue problem in
\eqref{eq:approx_stab_eq_2}

\subsection*{Statistics for networks with uniform parameters} \label{sec:strongly_conn}   

In order to gather statistics on the stability of closed unimolecular reaction networks, and to uncover the influence of
network-wide effects, we systematically generate (see Methods) and analyze all such reaction networks up to $K=6$
chemicals. To this end, we generated all \nscc strongly connected simple directed graphs containing up to $ K=6 $ nodes,
which includes 1, 5, 83, 5,048, and 1,047,008 graphs with $K=2$, 3, 4, 5, and 6 nodes, respectively \cite{oiesA035512}.
For examples of all reaction networks with $K=3$ and $K=4$ chemicals, see Figures S1 and S2 in the Supporting
Information. As described above, with the simplifying assumption of uniform parameters, the stability of a reaction
network is given by \eqref{eq:approx_stab_eq_2}, and depends only on the network topology as well as the value of the
dimensionless mobilities $ \mus $ and $ \mup $. We first study the stability of all the generated networks under eight
choices of the mobilities: $ (\mus, \mup) \in \left\{ (\pm 1, 0), (0, \pm 1), (\pm 1, \pm 1), (\pm 1, \mp 1) \right\} $.
In the following, we call these ``mobility patterns'', and for conciseness write $+$ or $-$ instead of $+1$ or $-1$.

\begin{table*}[hbtp] \centering \input{./tables/tab1_stabStatistics.tex} \caption{ Statistics for the stability of
        closed unimolecular reaction networks up to $K=6$ chemicals. Mobility patterns are written as ($\mus, \mup$).
        For each mobility pattern, the proportion of unstable reaction networks is given out of all \nscc networks,
        whereas the total is given out of all 8$\times$\nscc combinations of mobility patterns and reaction networks,
        i.e.~interaction networks. An interaction network is considered self-attracting if it verifies$ \sum_{m=1}^M
    \eta_{m,m} < 0 $. } \label{tab:stab_stats}
\end{table*}

A choice of reaction network and mobility pattern defines an interaction network (\cref{fig:act_cycles}\textit{D}). For
an example of all interaction networks with $K=3$ chemicals, see Figure S1 in the Supporting Information. We
systematically solve the eigenvalue problem in \eqref{eq:approx_stab_eq_2} for all $8 \times \nscc$ interaction
networks, and in each case calculate $\beta_\mathrm{max}$. This procedure yields values of $\beta_\mathrm{max}$ with
magnitudes spanning nine decades. Due to this wide range, we perform the corresponding calculations symbolically, in
order to distinguish small but non-null eigenvalues from null eigenvalues. Interaction networks with
$\beta_\mathrm{max}>0$ are labeled unstable.

As a point of comparison, we consider the special case of reactions in the saturated regime, $v_r (\bm{\rho}) = \alpha
\ccat{e_r}$, which leads to D-P interactions without network-wide effects, as previously described. In this special
case, the stability of the system is fully determined by the eigenvalue problem $\beta^\mathrm{sat}   \pertcat = - 
\tilde{\bm{\eta}}^\mathrm{sat} \pertcat$ where  $\tilde{\bm{\eta}}^\mathrm{sat} \equiv (\bm{\mu}/\mu_0)\bm{S}$ is a
dimensionless interaction matrix that depends only on $\mus$, $\mup$, and the topology of the reaction network. As in
the case of linear kinetics, we may define
$\beta_\mathrm{max}^\mathrm{sat}\equiv\text{max}(\text{Re}(\beta^\mathrm{sat}))$, and classify networks with
$\beta_\mathrm{max}^\mathrm{sat}>0$ as networks which are unstable in the case of saturated  kinetics.

In this way we obtain the stability statistics broken down by mobility pattern shown in \Cref{tab:stab_stats} (see Table
S1 in the Supporting Information for the statistics broken down by network size).
A first key observation is that slightly over $ 1/2$ %50\%  
of interaction networks are unstable for linear kinetics, compared to exactly $3/8$ for saturated kinetics, implying
that network-wide effects facilitate self-organization. We also calculate, for each mobility pattern, the proportion of
unstable networks which involve an overall self-attracting set of enzymes, i.e.~whose interaction matrix $
\tilde{\bm{\eta}} $ has a negative trace $ \sum_m \tilde{\eta}_{m,m} < 0 $ (and similarly with
$\tilde{\bm{\eta}}^\mathrm{sat}$ for saturated kinetics). Previous work has shown that, in the regime of saturated
kinetics, overall self-attraction is a necessary and sufficient condition for self-organization for enzymes interacting
through the production or consumption of a single chemical \cite{agudo-canalejo2019Active} and for enzymes participating
in a metabolic cycle if they have uniform parameters  \cite{ouazan-reboul2023Selforganization}, but not if their
parameters are not uniform \cite{ouazan2023network,ouazan-reboul2023Interaction}. We find that, for saturated kinetics,
an interaction network is unstable if and only if it satisfies this self-attraction condition. On the other hand, for
linear kinetics, we find that self-attraction is a sufficient, but not a necessary, condition: the network-wide effects
induced by the screening factor can overcome self-repulsion and lead to self-organization. Additionally, we find that
network-wide effects can make a network overall self-attracting, even if it would not be so for saturated kinetics.

\begin{figure}[t]
    \centering
    \includegraphics[width = \columnwidth]{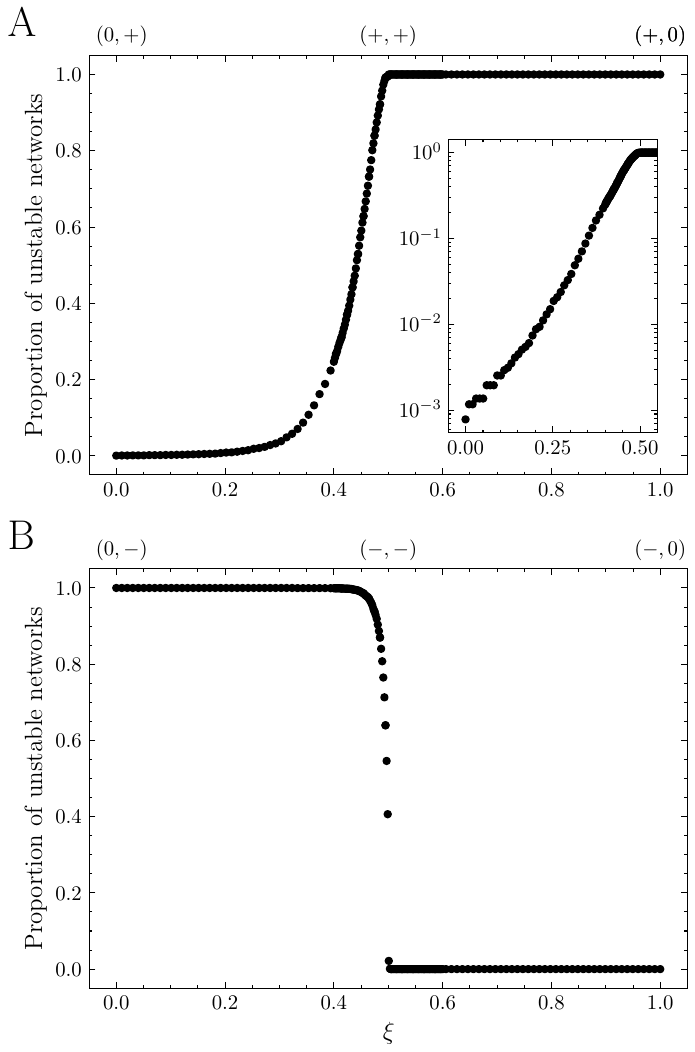}
    \caption{
        Transitions in the proportion of unstable networks under continuous changes in the mobility patterns, for all
        \nscconetofive closed unimolecular reaction networks up to $K=5$ chemicals and linear reaction kinetics. 
        (A) $ \left( \mus, \mup \right) = \left( \xi, 1 - \xi \right) $, and 
        (B) $ \left( \mus, \mup \right) = - \left( \xi, 1 - \xi \right) $. 
        Inset in (A): magnification showing the  exponential increase in the proportion of unstable networks (notice the
        logarithmic vertical axis).
    } \label{fig:cont_mob_change}
\end{figure}

Focusing on individual mobility patterns, we first find that purely substrate-chemotactic enzymes [pattern $(-, 0)$],
exclusively lead to stable networks for both linear and saturated kinetics. Qualitatively, this universal stability is
related to the fact that such enzymes are necessarily self-repelling, as they consume the chemical that they are
attracted by. Conversely, purely substrate-antichemotactic enzymes [pattern $(+, 0)$] consume a chemical by which they
are repelled, and are thus self-attracting and unstable for both types of reaction kinetics. Analogous qualitative
arguments based on self-interaction can explain the behavior of purely product-chemotactic enzymes [pattern $(0, -) $],
and the behavior of the antisymmetric patterns $ (-,+) $ and $ (+,-) $.

Surprisingly, in the case of product-antichemotactic enzymes [pattern $(0, +) $], a small proportion of networks, none
of which are self-attracting, is unstable only under linear kinetics. The presence of self-repelling and unstable
networks demonstrates that network-wide effects induced by the substrate concentration dependence of the reaction
kinetics can provide an alternate route towards self-organization. This mechanism for self-organization without
self-attraction is distinct from the one found in Refs.~\citenum{ouazan2023network,ouazan-reboul2023Interaction}, which
was based on saturated reaction kinetics, but non-identical activities and mobilities for the enzymes. Note that, since
all networks are unstable for pattern $(0, -) $, for the small subset of networks which is unstable for pattern $(0, +)
$  an instability can occur both if the enzymes are attracted to or repelled from their product.

For networks with symmetric mobility patterns $(-, -)$, enzymes do not self-interact under saturated kinetics, as
in this case the self-repulsion arising from substrate chemotaxis balances the self-attraction arising from product
chemotaxis. For this mobility pattern, we find that network-wide effects can create attractive self-interactions, which
constitutes yet another mechanism through which network-wide effects enable self-organization. This new mechanism
becomes especially prevalent and impactful in the other symmetric mobility pattern, $(+, +)$, for which all but a
handful of networks are unstable and self-attracting for linear kinetics, even if all networks are stable for saturated
kinetics. Again in this case, we find that there are networks that are unstable under both mobility patterns $(+, +)$
and $(-, -)$, implying that these reaction networks can show an instability both if the enzymes are attracted to or
repelled from their substrate and product.

In order to characterize the transition between stability and instability in the case of linear kinetics where
network-wide effects are important, we measure the proportion of unstable reaction networks under a continuous change of
the mobilities between each pair of previously studied mobility patterns. The changes in mobility patterns from $ (0, +)
$ to $ (+, 0) $ and from $ (0, -) $ to $ (-, 0) $ are of particular interest, as both cases involve a transition from
(almost) all reaction networks being stable to all being unstable. We continuously vary the mobilities according to a
parameter $ \xi \in \left\{ 0, 1 \right\}$ by imposing
$ \left( \mus, \mup \right) = \left(
\xi, 1 - \xi \right) $ (\cref{fig:cont_mob_change}A) or  $ \left( \mus, \mup \right) = -\left( \xi,  1 - \xi
\right) $ (\cref{fig:cont_mob_change}\textit{B}). In both cases,
$ \xi = 0.5 $ corresponds to a symmetric pattern, respectively $ (+, +) $ and $ (-, -) $.
Due to the large number of calculations involved, we limited this study to the \nscconetofive  reaction
networks containing up to $ K=5 $ chemical species.
We find that, for the transition between $ (0,+) $ and $ (+, 0) $, the proportion of unstable networks undergoes an
exponential (\cref{fig:cont_mob_change}\textit{A}, inset) increase from $ 0 $ to $ 1 $ between $ \xi = 0 $ and $ 0.5 $.
For the transition between $ (0,-) $ and $ (-,0) $, on the other hand, we observe an abrupt change from all networks
being stable to all unstable around $ \xi \approx 0.5  $.

\begin{figure*}[hbtp]
    \centering
    \includegraphics[width = 0.99\textwidth]{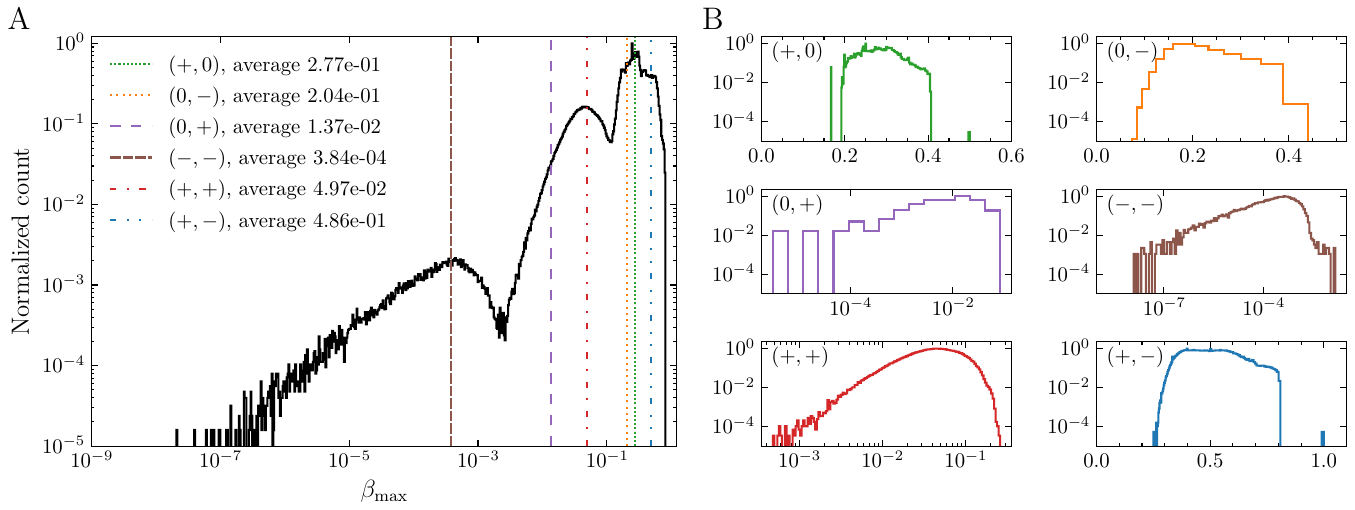}
    \caption{
        Distribution of $\beta_\mathrm{max}>0$ for all the unstable closed unimolecular reaction networks up to $K=6$ chemicals that are unstable and all six
         mobility patterns that allow for instabilities.
        (A) Distribution over all mobility patterns, with the average $\beta_\mathrm{max}$ for each mobility pattern shown as a dashed vertical line.
        (B) Distribution for each of the six unstable mobility patterns. Note the different range and the use of linear and logarithmic scales for the x-axis in each case.
    } \label{fig:eigen_density}
\end{figure*}

\subsection*{Distribution of growth rates for networks with uniform parameters} \label{sec:instab_rates}
To further quantify the degree to which different interaction networks are unstable, we computed the probability
distributions of $\beta_\mathrm{max}$ for unstable interaction networks (i.e.~confined to values
$\beta_\mathrm{max}>0$). Indeed, through the instability condition in \cref{eq:beta_inst}, larger values of
$\beta_\mathrm{max}$ correspond to wider ranges of values of $\Dps$, $\mu_0$, and $c_\mathrm{tot}$ that can induce an
instability. In this analysis, we considered all reaction networks with up to $K=6$ chemicals, and the six mobility
patterns that can become unstable (\Cref{tab:stab_stats}). When pooling together all six mobility patterns, the
resulting distribution is fat-tailed (\cref{fig:eigen_density}\textit{A}), spanning the range  $ \beta_\mathrm{max} \in
\left(10^{-9}, 1 \right) $,  with an overall average of $ \left< \beta_\mathrm{max}  \right> = 2.527 \cdot 10^{-1}$.

The wide range of $\beta_\mathrm{max}$-values observed can in part be explained by considering the distributions
associated with individual mobility patterns (\cref{fig:eigen_density}\textit{B}). Each pattern displays a markedly
different distribution, with mean values and typical ranges that are consistent with qualitative arguments based on the
presence of self-attraction in the interaction network. For the mobility patterns $ (+,0)$, $ (0,-)$, and $ (+,-) $,
which are associated with self-attracting species for both linear and saturated reaction kinetics, the 
$\beta_\mathrm{max}$-values are all on the order of $ 10^{-1}$, and have a mean value on the order of the average of the
overall distribution. In particular, patterns $ (+,0) $ and $ (0,-) $, which correspond to similarly-strong degrees of
self-attraction, broadly span the same range of  $\beta_\mathrm{max}$-values. The pattern $ (+,-) $, on the other hand,
makes the enzymes the most self-attracting, as it corresponds to species both repelled by the chemical they consume and
attracted to the one they produce. As a consequence, its  $\beta_\mathrm{max}$-distribution features a markedly larger
mean and contains the overall largest values.

We now focus on the mobility patterns $ (-,-)$, $(0,+)$, and $(+,+)$, which lead to self-organization only when
network-wide effects are present. The associated $\beta_\mathrm{max}$-distributions have means that are several orders
of magnitude smaller than the previously discussed patterns, and span wider ranges (typically several decades). The
pattern $ (+,+) $, for which network-wide effects induce self-attraction, displays significantly larger
$\beta_\mathrm{max}$-values than $ (-,-) $ and $ (0,+) $, for which network-wide effects enable self-organization in
spite of self-repulsion. From these statistics we deduce that, while network-wide effects can indeed favor instabilities
in systems without D-P self-attraction, the instabilities in these systems will be harder to observe, because their
smaller $\beta_\mathrm{max}$-values require either significantly smaller values of the enzyme diffusion coefficient
$\Dps$, larger values of their chemotactic mobility $\mu_0$, or larger total amounts of reactant $c_\mathrm{tot}$, as
seen from the instability condition in \cref{eq:beta_inst}. This is particularly true for interaction networks which
stay self-repelling even in the presence of network-wide effects: such networks are rarely unstable, and only feature
very small $\beta_\mathrm{max}$-values. We also speculate that the wider distribution of $\beta_\mathrm{max}$  for these
mobility patterns is due to an increased sensitivity to the network structure, resulting from the importance of
network-wide effects for such patterns.

\section*{Discussion}
In this work, we have shown that the spatial dynamics of the components of a catalyzed chemical reaction network
(including diffusion for the chemicals, and diffusion as well chemotaxis in response to chemical gradients for the
enzymes) can lead to self-organization of the enzymes through a spatial instability. We first characterized this
instability process by applying linear stability analysis on generic network structures and chemical reaction rates,
obtaining a concise criterion for the observation of spatial instabilities in such systems, see
Eq.~\eqref{eq:approx_stab_eq}. From the structure of the effective enzyme-enzyme interaction matrix involved in this
stability equation, we deduced that perturbations of the enzyme concentration fields are effectively propagated through
the chemical reaction network, giving rise to network-wide effects. In particular, two enzymes can effectively interact
with each other even if they do not undergo chemotaxis towards each other's substrates or products. We then focused on
unimolecular reaction networks with linear reaction rates. We systematically determined the stability of all strongly
connected reaction networks involving six or fewer chemical species for various patterns of chemotactic mobilities,
finding that the aforementioned network-wide effects can facilitate self-organization. By interpolating between
different mobility values, we discovered that continuous changes in mobilities can either lead to an exponential
increase of the number of unstable networks, or to a sharp transition. We finally quantified the extent to which
different mobility patterns are unstable which, due to network-wide effects, can span nine orders of magnitude.

In order to gather gain an initial understanding of self-organization processes in catalyzed reaction networks, we
limited most of our quantitative exploration to small, closed, unimolecular reaction networks. Real-life biochemical
reaction networks, however, are typically larger and more complex than the ones we studied, and involve bimolecular
reactions as well as external inputs and outputs of reaction material. Our general framework culminating in
Eq.~\eqref{eq:approx_stab_eq}, however, can be directly applied to bimolecular reactions, and easily extended to account
for non-enzymatic chemical sources or sinks such as chemostats. A possible approach for the analysis of large realistic
networks could be to use small, strongly connected networks as building blocks for more complex topologies closer to
metabolic pathways, akin to the network motifs formalism used in systems biology
\cite{alon2007Network,milo2002Network,ouazan-reboul2023Interaction}. Another relevant extension of this work is to
determine how enzyme self-organization influences the output of biochemical pathways. Concentrating enzymes into
condensates has indeed been shown to result in localized metabolic factories in which pathway outputs are increased
through the channeling of reaction intermediates between enzymes \cite{zhang2021Metabolons,harmon2022Molecular}, or to
favor one branch of a pathway with respect to another \cite{hinzpeter2019Regulation}. The effect of spatial structure on
the output of biochemical pathways has so far mostly been studied by starting from a pre-defined spatial structure.
However, recent works have shown that a single enzyme catalyzing a one-step reaction can undergo phase separation,
lowering its activity in the process \cite{cotton2022CatalysisInduced}. This work is a first step towards bridging these
two approaches. It will be of particular interest to determine which kinds of pathway regulations are possible through
phase separation in multi-enzyme systems, and how spontaneously-arising spatial structure affects metabolic output.

\section*{Materials and methods}

\subsection*{Generation of strongly connected reaction networks}
Above, we systematically analyzed the stability of minimal closed networks involving unimolecular reactions.
To do so, we generated the adjacency matrices corresponding to all \nscc strongly connected digraphs with $ K=2$ to 6 nodes, where the nodes take the role of the chemicals involved in the reactions.
This was done by first using the command line graph isomorphism program nauty \cite{mckay2014Practical}, which
can generate one component of each graph equivalence class with a certain number of nodes, yielding the set of all
unique digraphs with a given size.
We then applied a Julia implementation of Tarjan's algorithm \cite{tarjan1972DepthFirst} to each obtained digraph, which
returned the set of strongly connected components for each digraph.
Only digraphs with a single strongly connected component (which are by definition strongly connected) are
then kept.
We verified that the number of digraphs obtained with this method is consistent with the value reported in literature,
obtained from  \cite{oiesA035512}, and that the adjacency matrices obtained for $ K=3 $ and 4 nodes correspond to the
recorded examples of strongly connected digraphs.

\section*{Acknowledgements}
We acknowledge support from the Max Planck School Matter to Life and the MaxSynBio Consortium which are jointly funded
by the Federal Ministry of Education and Research (BMBF) of Germany and the Max Planck Society.

% Bibliography
\bibliographystyle{apsrev4-2}
\bibliography{networks,Golestanian}

\clearpage
\onecolumngrid
\appendix
\section*{Supplemental Information}
\setcounter{section}{0}
\renewcommand{\thesection}{S\arabic{section}}
\renewcommand{\thefigure}{S\arabic{figure}}
\renewcommand{\thetable}{S\arabic{table}}
\renewcommand{\theequation}{S\arabic{equation}}
\setcounter{figure}{0}
\setcounter{table}{0}
\setcounter{equation}{0}

\FloatBarrier

\begin{table}
    \centering
    \include{tables/SItab1_stabStatisticsBySize}
    \caption{Stability statistics broken down by number of chemicals.}
\end{table}

\begin{figure}
    \centering
    \includegraphics[width=0.75\textwidth]{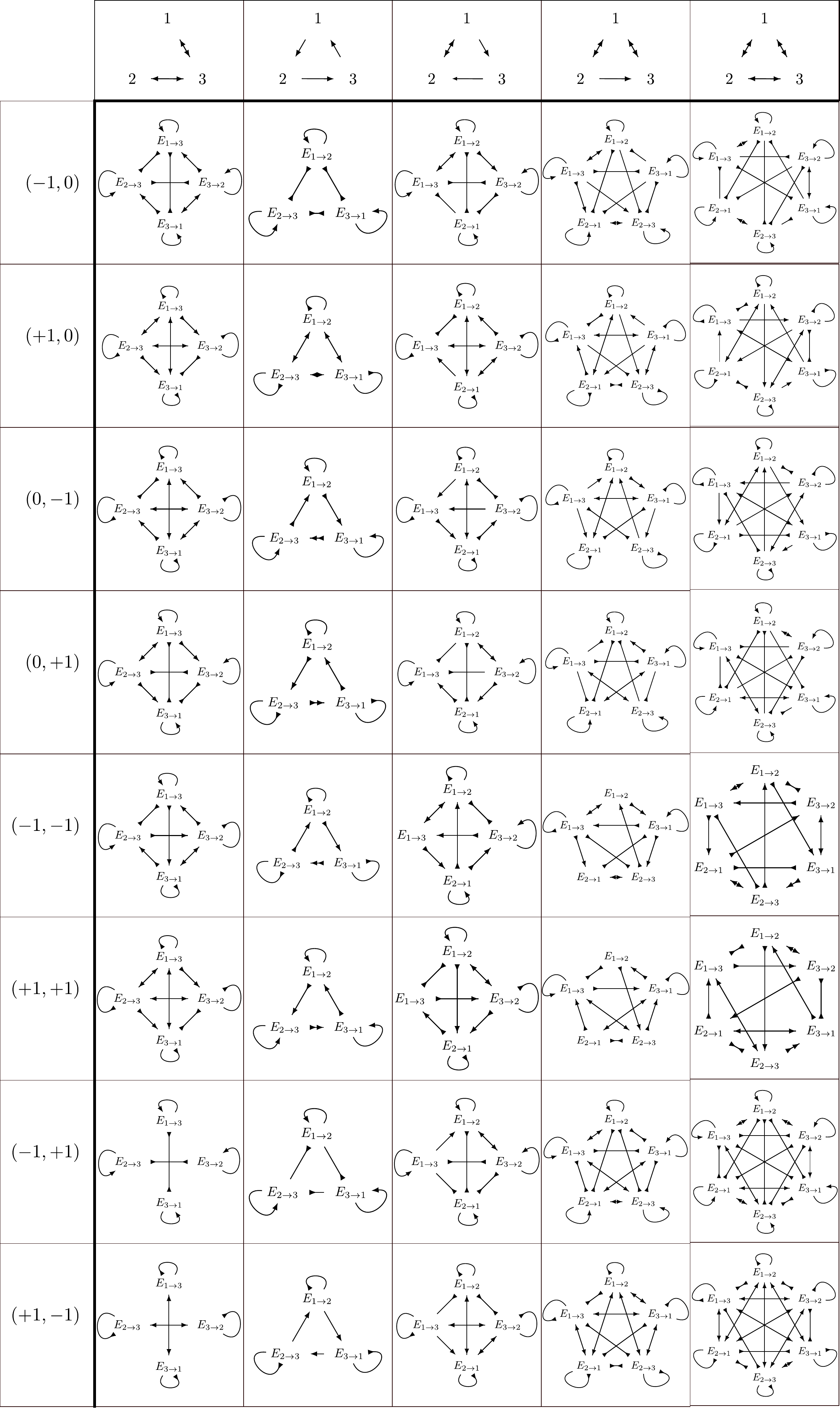}
    \label{supfig:size_3_int_nets}
    \caption{All $8 \times 5=40$ interaction networks for $K=3$ chemical species.
    Top row represents the catalyzed reaction networks between chemicals indexed 1, 2, and 3 in the form of digraphs, where a directed link between chemicals
    $i$ and $j$ represents the existence a catalyzed reaction converting $i$ into $j$.
    Leftmost column corresponds to the relative substrate and product mobilities $(\mu_\mathrm{s}, \mu_\mathrm{p})$.
    In the main portion of the table, catalyzed reaction networks between enzymes $E_{i \rightarrow j}$ catalyzing the conversion of chemical $i$ into chemical
    $j$ are shown.
    The convention for attracting and repulsing interactions is the same as in the main text.
    }
\end{figure}

\begin{figure}
    \centering
    \includegraphics[width=0.9\textwidth]{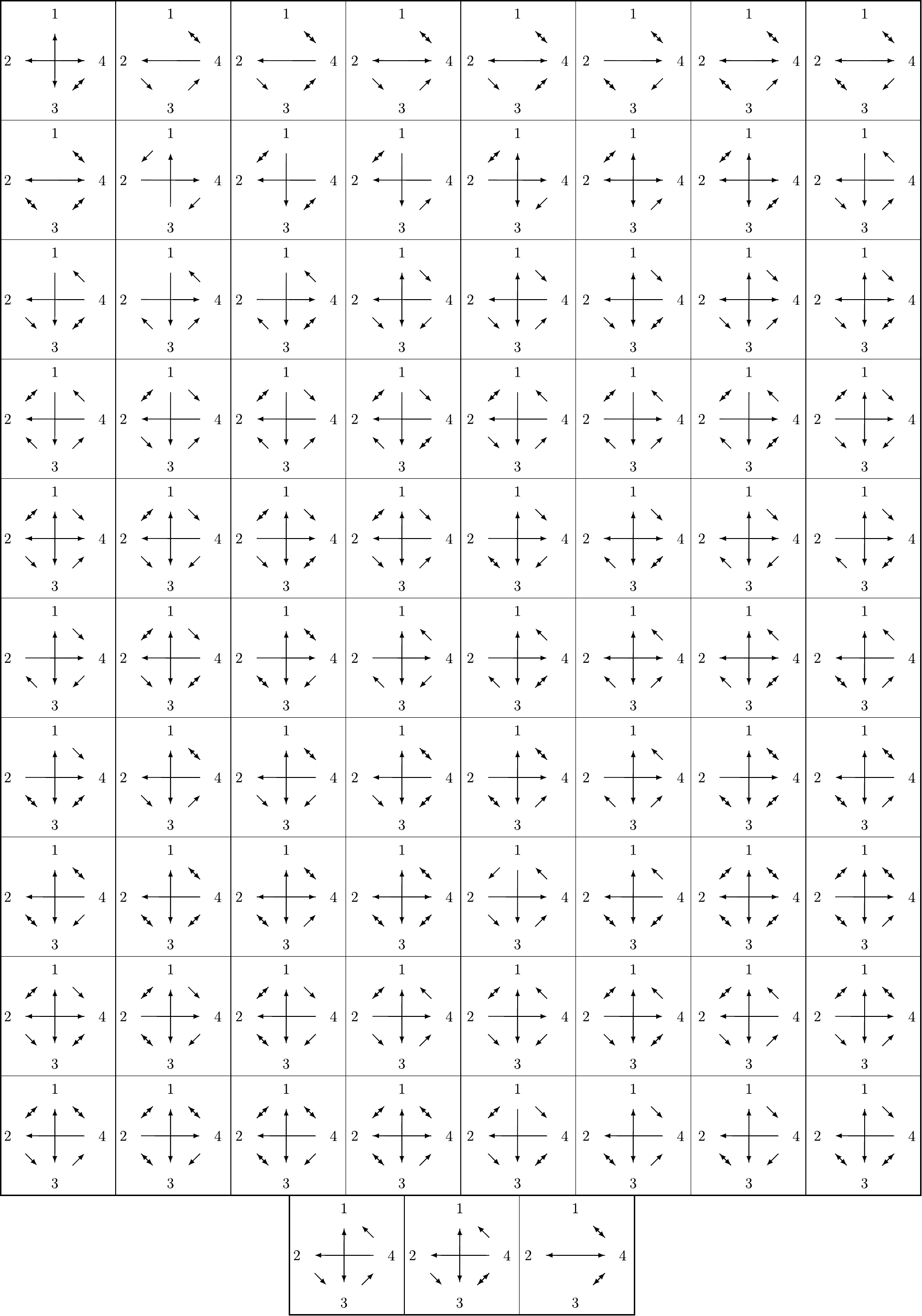}
    \caption{All 83 reaction networks for $K=4$ chemical species. The representation of the chemical networks is the same as in Fig. S1, top row.}
\end{figure}
\end{document}

%% file: tables/tab1_stabStatistics.tex
% See data below.
% Taken from: "/Users/vincent/research/projects/19_selfOrgCatActMix/240823_revisitingNetworks_essentialsOnly/23_02-standardizedAnalysis/Eigvals_statistics-Copy1.ipynb"
$
\begin{array}{ccccc}
\toprule
 & \multicolumn{2}{c}{\text{Linear kinetics (Full interactions)}} & \multicolumn{2}{c}{\text{Saturated kinetics (Naive interactions)}}\\
\midrule
\text{Mobilities} & \text{Proportion unstable} & \text{Of which self-attr.} & \text{Proportion unstable} & \text{Of which self-attr.} \\
\midrule
% ---------- Substrate-responding ---------- %
\begin{array}{c}
(-, 0) \\
(+, 0) \\
\end{array} & 
\begin{array}{c}
0 \\
1 
\end{array}  &
\begin{array}{c}
\text{N/A} \\
1 
\end{array} &
\begin{array}{c}
0 \\
1
\end{array} & 
\begin{array}{c}
\text{N/A} \\
1
\end{array}
\\ \addlinespace
% ---------- Product-responding  ---------- %
\begin{array}{c}
(0, -) \\
(0, +) \\
\end{array} & 
  \begin{array}{c}
1 \\
 2.167 \cdot 10^{-4} 
\end{array}
&
\begin{array}{c}
1 \\
0 
\end{array} &
\begin{array}{c}
1 \\
0 
\end{array} &
 \begin{array}{c}
1 \\
\text{N/A} 
\end{array}
\\ \addlinespace
% ---------- Symmetric  ---------- %
\begin{array}{c}
(-, -) \\
(+, +) \\
\end{array} & 
 \begin{array}{c}
         2.405 \cdot 10^{-2}\\
    1 - 1.388 \cdot 10^{-4}
\end{array} 
&
\begin{array}{c}
1.976 \cdot 10^{-4}\\
1 - 4.753 \cdot 10^{-6}
\end{array} &
 \begin{array}{c}
0 \\
0 
\end{array} &
 \begin{array}{c}
\text{N/A} \\
\text{N/A}
\end{array}
\\ \addlinespace
% ---------- Antisymmetric  ---------- %
\begin{array}{c}
(-, +) \\
(+, -) \\
\end{array} & 
 \begin{array}{c}
0 \\
1
\end{array} 
&
\begin{array}{c}
\text{N/A} \\
1
\end{array} & 
 \begin{array}{c}
0 \\
1
\end{array} & 
 \begin{array}{c}
\text{N/A} \\
1 
\end{array}
\\ \addlinespace
\text{Total:} & 0.5030  & 0.9939 & 0.3750 & 1 \\
\bottomrule
\end{array}
$
%Unstable row proportions:
%["0.000e+00" "1.000e+00" "1.000e+00" "2.167e-04" "2.405e-02" "9.999e-01" "0.000e+00" "1.000e+00"]
%Written as 1-x
%["1.000e+00" "0.000e+00" "0.000e+00" "9.998e-01" "9.760e-01" "1.388e-04" "1.000e+00" "0.000e+00"]
%Unstable proportions for pairs of mobilities:

%1×4 Matrix{String}:
 %"0.000e+00"  "2.167e-04"  "2.405e-02"  "0.000e+00"

%Written as 1-x
%["1.000e+00" "9.998e-01" "9.760e-01" "1.000e+00"]
%Total unstable rate
%5.030e-01

%Proportion of unstable networks which are self-attracting
%["NaN" "1.000000e+00" "1.000000e+00" "0.000000e+00" "1.976128e-04" "9.999952e-01" "NaN" "1.000000e+00"]
%Written as 1-x
%["NaN" "0.000000e+00" "0.000000e+00" "1.000000e+00" "9.998024e-01" "4.752856e-06" "NaN" "0.000000e+00"]
%Total proportion of usntable networks which are self-attracting

%"9.939702e-01"

%% file: tables/SItab1_stabStatisticsBySize.tex
$
\begin{array}{ccccccc}
\toprule
 & \multicolumn{6}{c}{\text{Proportion of unstable networks with} \ K \
\text{chemicals}}\\
\midrule
\text{Mobilities}& K = 2 & K = 3 & K = 4 & K = 5 & K = 6 & K \leq 6\\
\midrule
% ---------- Substrate-responding ---------- %
\begin{array}{c}
(-, 0) \\
(+, 0) \\
\end{array} & 
\begin{array}{c}
0 \\
1 
\end{array}  &
\begin{array}{c}
0 \\
1 
\end{array}  &
\begin{array}{c}
0 \\
1 
\end{array}  &
\begin{array}{c}
0 \\
1 
\end{array}  &
\begin{array}{c}
0 \\
1 
\end{array}  &
\begin{array}{c}
0 \\
1 
\end{array} 
\\ \addlinespace
% ---------- Product-responding  ---------- %
\begin{array}{c}
(0, -) \\
(0, +) \\
\end{array} & 
\begin{array}{c}
1 \\
0
\end{array}&
\begin{array}{c}
1 \\
0
\end{array}&
\begin{array}{c}
1 \\
0
\end{array}&
  \begin{array}{c}
     1\\
     7.924 \cdot 10^{-4}
\end{array} &
 \begin{array}{c}
     1\\
     2.139 \cdot 10^{-4}
\end{array} &
 \begin{array}{c}
1 \\
 2.167 \cdot 10^{-4} 
\end{array}  
\\ \addlinespace
% ---------- Symmetric  ---------- %
\begin{array}{c}
(-, -) \\
(+, +) \\
\end{array} & 
\begin{array}{c}
    0\\
    0
\end{array}&
 \begin{array}{c}
    0\\
    0.6
\end{array}&
 \begin{array}{c}
    0\\
    9.277 \cdot 10^{-1}
\end{array}&
\begin{array}{c}
    8.518 \cdot 10^{-2}\\
    9.958 \cdot 10^{-1}
\end{array} &
\begin{array}{c}
    2.376 \cdot 10^{-2}\\
    1- 1.108 \cdot 10^{-4}
\end{array} &
\begin{array}{c}
    2.405 \cdot 10^{-2}\\
    1 - 1.388 \cdot 10^{-4}
\end{array} 
\\ \addlinespace
% ---------- Antisymmetric  ---------- %
\begin{array}{c}
(-, +) \\
(+, -) \\
\end{array} & 
 \begin{array}{c}
0 \\
1
\end{array}&
\begin{array}{c}
0 \\
1
\end{array}&
 \begin{array}{c}
0 \\
1
\end{array}&
 \begin{array}{c}
0 \\
1
\end{array}&
 \begin{array}{c}
0 \\
1
\end{array}&
 \begin{array}{c}
0 \\
1
\end{array}
\\ \addlinespace
    \text{Total:} & 0.375 & 0.45 & 0.491 & 0.5102 & 0.5030 & 0.5030\\
\bottomrule
\end{array}
$